\documentclass[
	aps,
	showpacs,
	superscriptaddress,
	preprintnumbers,
	amsmath,
	twoside,
	amsfonts,
	twocolumn,
	amssymb,
	prb]		
{revtex4-1}

\usepackage{bibunits}		
\usepackage{graphicx}
\usepackage{hyperref}		
\usepackage{color}		
\usepackage{amsmath}
\usepackage{amsfonts}
\usepackage{amssymb}
\usepackage{float}		
\usepackage{lmodern}		

\def \be {\begin{equation}}
\def \being {\begin{equation*}}
\def \en {\end{equation}}
\def \ening {\end{equation*}}

\def \d {^\dagger}

\def \action {\mathcal{S}}
\def \veck {\mathbf{k}}
\def \veckp {\mathbf{k}'}
\def \vecq {\mathbf{q}}
\def \vecr {\mathbf{r}}
\def \sites {N}
\def \flavours {\mathcal{N}}

\def \ckxcky {\cos k_x^{} \cos k_y^{}}
\def \ckxpcky {\cos k_x^{} + \cos k_y^{}}
\def \ckxmcky {\cos k_x^{} - \cos k_y^{}}

\usepackage{graphicx}
\usepackage{dcolumn}
\usepackage{bm}
\usepackage{hyperref}
\usepackage{multirow}
\usepackage{array}
\usepackage{booktabs}
\usepackage{upgreek}
\usepackage{epsfig}
\usepackage{mathrsfs}
\usepackage{amssymb}
\usepackage{amsbsy}
\usepackage{color}
\usepackage{cancel}
\usepackage{pifont}
\usepackage{marginnote}
\usepackage{float}
\usepackage{tikz}
\usepackage{ctable}
\usetikzlibrary{arrows}
\tikzstyle{block}=[draw opacity=0.7,line width=1.4cm]
\usetikzlibrary{positioning}
\usepackage[caption=false]{subfig}
\usepackage{setspace}

\newcommand{\bcen}{\begin{center}}
\newcommand{\ecen}{\end{center}}
\newcommand{\btab}{\begin{tabular}}
\newcommand{\etab}{\end{tabular}}
\newcommand{\bdes}{\begin{description}}
\newcommand{\edes}{\end{description}}

\newcommand{\beq}{\begin{equation}}
\newcommand{\eeq}{\end{equation}}
\newcommand{\bea}{\begin{eqnarray}}
\newcommand{\eea}{\end{eqnarray}}

\newcommand{\bary}{\begin{array}}
\newcommand{\eary}{\end{array}}
\newcommand{\benum}{\begin{enumerate}}
\newcommand{\eenum}{\end{enumerate}}
\newcommand{\bitem}{\begin{itemize}}
\newcommand{\eitem}{\end{itemize}}

\newcommand{\bk} { \bm{k} }

\newcommand{\bzero} { {\boldsymbol{0}}}

\newcommand{\dou}{\partial}

\newcommand{\eqn}[1] {eqn.~(\ref{#1})}

\newcommand{\Fig}[1]{Fig.~\ref{#1}}

\makeatletter

\newcommand{\Rmnum}[1]{\expandafter\@slowromancap\romannumeral #1@}
\makeatother

\newcommand{\SP}{P_s}
\newcommand{\AP}{P_a}
\newcommand{\SA}{A_s}
\renewcommand{\AA}{A_a}
\newcommand{\SPM}{\SP \text{-mode}}
\newcommand{\APM}{\AP \text{-mode}}
\newcommand{\SAM}{\SA \text{-mode}}
\newcommand{\AAM}{\AA \text{-mode}}

\newcommand{\MAPM}{M_{\AP}}
\newcommand{\MSAM}{M_{\SA}}
\newcommand{\MAAM}{M_{\AA}}
\newcommand{\doping}{p}

\def \ckxcky {\cos k_x^{} \cos k_y^{}}
\def \ckxpcky {\cos k_x^{} + \cos k_y^{}}
\def \ckxmcky {\cos k_x^{} - \cos k_y^{}}

\newcommand{\ourtitle}{Crucial role of Internal Collective Modes in Underdoped Cuprates}

\begin{document}


\title{\ourtitle}
\author{Aabhaas V. Mallik}
\email{aabhaas@physics.iisc.ernet.in}
\affiliation{Department of Physics, Center for Condensed Matter Theory, Indian Institute of Science, Bengaluru - 560012, India}
\author{Umesh K. Yadav}
\email{umesh@physics.iisc.ernet.in}
\affiliation{Department of Physics, Center for Condensed Matter Theory, Indian Institute of Science, Bengaluru - 560012, India}
\author{Amal Medhi}
\email{amedhi@iisertvm.ac.in}
\affiliation{School of Physics, Indian Institute of Science Education and Research, Thiruvananthapuram - 695016, India}
\author{H. R. Krishnamurthy}
\email{hrkrish@physics.iisc.ernet.in}
\affiliation{Department of Physics, Center for Condensed Matter Theory, Indian Institute of Science, Bengaluru - 560012, India}
\author{Vijay B. Shenoy}
\email{shenoy@physics.iisc.ernet.in}
\affiliation{Department of Physics, Center for Condensed Matter Theory, Indian Institute of Science, Bengaluru - 560012, India}
\date{\today}

\begin{abstract}
The enigmatic cuprate superconductors have attracted resurgent interest with several recent reports and discussions of competing orders in the underdoped side. Motivated by this, here we address the natural question of fragility of the $d$-wave superconducting state in underdoped cuprates. Using a combination of theoretical approaches we study $t$-$J$ like models, and discover an -- as yet unexplored --  instability that is brought about by an ``internal'' (anti-symmetric mode) fluctuation of the $d$-wave state. This new theoretical result is in good agreement with recent STM and ARPES studies of cuprates. We also suggest experimental directions to uncover this physics.
\end{abstract}
\maketitle

Cuprate superconductors, apart from their obvious importance from the stand point of applications, provide some of the most puzzling and fascinating phenomena in physics\cite{OrensteinScience00}. In the process of their exploration, the highly unconventional nature of their phases has become evident\cite{ShenPNAS12,HoffmanScience14,DavisScience14}. While it is widely accepted that the underdoped superconducting state has small superfluid stiffness ($\rho_s^{}$)\cite{Kotliar2PRB88,KivelsonNature95}, the presence of competing or intertwined orders has also often been reported\cite{KimNatLett10,TaoNature11} and theorized\cite{FradkinRMP15,BalentsPRB05,LaughlinPRL14,TroyerPRL14}. The associated pseudogap phase has remained a major open problem\cite{NormanAdvPhys05,TailleferNatLett10,CampuzanoPNAS11}. A significant fraction of the literature considers it to be a consequence of strong pairing and comparatively small $\rho_s^{}$ in the underlying $d$-wave superconducting ($d$-SC) ground state\cite{SawatzkyRPP08,Dasgupta2PRB11,CampuzanoNatLett14}. More recently, the view that the pseudogap regime cannot be understood entirely on the basis of superconducting fluctuations and the role of competing orders may be crucial\cite{DemsarPRL99,KondoNatPhys10} has been gaining ground. 

The status described above raises some natural questions, e.g., does the ``fragility'' of the $d$-SC state on the underdoped side arise from the small $\rho_s^{}$, or are there other reasons? The role of other collective modes\cite{MandalPRB00,LeePRB03} about the $d$-SC state has got little attention\cite{LiuNatCommun16}. This motivates our investigation of the fluctuations of the $d$-SC state seeking possible additional causes of its fragility. Moreover, there have been recent advances in the study of Higgs (amplitude) mode fluctuations in superconductors\cite{MatsunagaPRL13,ShermanNatPhys15}; our study is useful in that context as well.

The collective excitations of the $d$-SC state relate to the long wavelength fluctuations of the phase and amplitude of the superconducting pairing field. The $d$-SC state on a square lattice has a pairing field whose value on the  $y$-bond has a $\pi$-phase relative to that on the $x$-bond in the same unit cell. This leads to four types of collective modes. The first two are the ``symmetric'' modes where the pairing field on both $x$ and $y$ bonds attached to a unit cell fluctuate in phase, and thus, preserve the local $d$-wave structure of the pairing field. The second type are the ``anti-symmetric'' modes where the fluctuations on the $x$ and $y$ bonds of a unit cell are of {\em opposite} sign with respect to each other (akin to an ``optical phonon'' or ``internal mode''), and thus, do not preserve the $d$-wave nature of the pairing field. The symmetric variety consists of the symmetric phase mode ($\SPM$), in which the fluctuation is in the phase of the pairing field, and is the same on both $x$ and $y$ bonds of a unit cell (fluctuation varies from one unit cell to another), and a similarly defined $\SAM$ where the amplitude of the pairing field fluctuates. On the other hand, the anti-symmetric phase mode ($\APM$), has a fluctuation in the phase of the pairing field which is opposite in sign on the $x$ and $y$ bonds of a unit-cell, and likewise for the anti-symmetric amplitude mode ($\AAM$), where the amplitude of the pairing field fluctuates (see illustrations in SM S1). In a stable $d$-wave superfluid induced by finite ranged interactions, the $\SPM$ corresponds to the gapless Goldstone mode, while the other modes are all gapped with gap parameters\footnote{The low energy action for a gapped mode $\phi$ is of the form, $\phi^{*}_{}(q) \left[ a - b(iq_l^{})^2_{} + c|\mathbf{q}|^2_{} + d\ q_x^{} q_y^{} \right] \phi(q)$. Here, $a/b$ is the gap of the mode. We call $a$ as the gap parameter, and find it relevant for our discussions.} $\MSAM$, $\MAPM$ and $\MAAM$. In real materials, e.g.. cuprates, the nature of the $\SPM$ is modified significantly due to its coupling to electromagnetic fields, but the $\APM$ remains largely unaffected (see SM S4).

In this paper we study the properties of these collective excitations of the $d$-SC state in $tJ$-like models\cite{RicePRL88,FukuyamaRPP08}, appropriate for cuprate superconductors, as a function of the hole doping $\doping$ using a number of methods, including functional techniques, large-$\flavours$ approximation, and numerical variational Monte Carlo method. A crucial finding of this work is the vanishing of the gap parameter of the {\em anti-symmetric phase ($\AP$) mode} at a finite hole doping $\doping_c$ ($\sim$ 0.06 for typical cuprate parameters) rendering the $d$-SC state unfavorable as a ground state for $\doping < \doping_c$. Furthermore, even in the regime $\doping \gtrsim \doping_c$, the gap parameters associated with the anti-symmetric internal modes $\AP$ and $\AA$ are much smaller than the other scales, pointing to the fragility of the $d$-SC state on the underdoped side. In particular, we show that these lead to a remarkable suppression of the $d$-wave pairing amplitude. This suggests that the experimentally observed pseudogap is likely to have physics beyond $d$-pairing. Our findings are in agreement with recent STM\cite{KimNatLett10,HoffmanScience14} and ARPES\cite{ShenPNAS12} studies on cuprate superconductors. We also discuss future experimental possibilities of verifying this physics.
 
\noindent
{\bf Model:} All our analytical results are obtained for the renormalized $t$-$J$ model\cite{ShibaSST88,AndersonJPCM04,LeeRMP06} on a 2D square lattice,
\begin{multline}
		\label{eq:H_Gutz}
		H = -g_t^{}(p) \sum_{\stackrel{i, \boldsymbol \delta}{\sigma}} t(\boldsymbol \delta) c_{i+\boldsymbol \delta \sigma} \d c_{i \sigma}^{}+g_s^{}(p)J\sum_{\langle i,j \rangle} \mathbf{S}_i \cdot \mathbf{S}_j\\
		- g_n^{}(p)J\sum_{\langle i,j \rangle} \frac{1}{4} n_i n_j - \mu \sum _{i} n_{i}
\end{multline}
where $c_{i \sigma}^{}$ ($c_{i \sigma} \d $) is the annihilation (creation) operator for an electron with $z$ component of spin $\sigma \in \{\uparrow,\downarrow\}$ at the $i$th site; $\mathbf{S}_i$ and $n_i^{}$ are, respectively, the fermion spin and occupation number operators at the $i$th site. $t(\boldsymbol \delta)$ is the hopping amplitude from any site $i$ to its neighbor at $i+\boldsymbol \delta$, while $J$ and $\mu$ are, respectively, the exchange interaction strength and chemical potential. $g_t^{}(p)$, $g_s^{}(p)$ and $g_n^{}(p)$ are doping ($\doping$) dependent Gutzwiller factors used to incorporate the effect of large $U$ projection while working with the full unprojected Hilbert space. The results discussed below are obtained using physically motivated choice of parameters\cite{FukuyamaRPP08} suitable for satisfactory description of cuprates; the next nearest neighbor hopping amplitude, $t'=-0.3t$, and the exchange interaction, $J = 0.3t$. For the Gutzwiller factors, we use $g_t^{}(p)=p$ and $g_s^{}(p)=g_n^{}(p)=1$\cite{BaskaranSSC87,Kotliar1PRB88}. We emphasize that the qualitative features of the results that we present below are insensitive to different reasonable choices of Gutzwiller factors. The temperature $T = 1/\beta$ is eventually set to zero. In the rest of this paper, we present the results and discuss their significance, and relegate the details of the calculations to the appended Supplemental Material (SM).

\noindent
\textbf{Saddle point and fluctuations:} To study the $d$-SC state we first obtain the action ($\action \left[ \{ \bar{c}_{k' \sigma'}^{}, c_{k' \sigma'}^{} \}\right]$) corresponding to the Matsubara-momentum space version of \eqn{eq:H_Gutz}. Next, we introduce Hubbard-Stratonovich fields $\Delta _{\alpha}^{} (q)$ and $K _\alpha ^{} (q)$, $\alpha \in \{ 0,1 \}$ or $\{ x,y \}$, to decompose the quartic terms in the action in the pairing and ``Fock'' channels, respectively. We refer to $\Delta _\alpha ^{} (q)$ as pairing fields, and $K _\alpha ^{} (q)$ as ``Fock'' fields. Now, on integrating out the quadratic fermionic fields, we obtain an effective action ($\tilde{\action} \left[ \{ \Delta_{\alpha'}^{}(q'), K_{\alpha'}^{}(q') \} \right]$) solely in terms of $\Delta _\alpha ^{} (q)$ and $K _\alpha ^{} (q)$. The saddle point of this action consistent with $d$-SC order ($\Delta_\alpha ^{} (q) = (-1)^\alpha_{} \sqrt{\sites \beta }\Delta_{SP}^{} \delta _{q,0}^{}$ and $K_\alpha ^{} (q) = \sqrt{\sites \beta }K_{SP}^{} \delta _{q,0}^{}$; $N$ is the number of lattice sites) is then found by solving the saddle point equations, $\delta \tilde{\action} / \delta \Delta^{*}_{\alpha}(q) = 0$ and $\delta \tilde{\action} / \delta K^{*}_{\alpha}(q) = 0$, along with the number equation, $\sites(1-p) = -\left( \partial F^{SP}_{} / \partial \mu \right) _{T=0}^{}$. Here $F^{SP}_{}$ is the saddle point grand free energy (see SM S1).

Fluctuations in the pairing field about this saddle point are investigated by setting $\Delta_\alpha ^{} (q) = (-1)^\alpha_{} \sqrt{\sites \beta }\Delta_{SP}^{} \delta _{q,0}^{} + \eta _\alpha ^{} (q)$ and $K_\alpha ^{}(q) = \sqrt{\sites \beta }K_{SP}^{} \delta _{q,0}^{}$ in the effective action, and expanding the effective action to quadratic power in the pairing fluctuation fields $\eta^{}_\alpha(q)$.
	\be
	\label{eq:action_gaussian}
	\tilde{\action} \left[ \{ \eta _{\alpha'} ^{*} (q'), \eta _{\alpha'} ^{} (q')\} \right] \simeq \tilde{\action} ^{SP}_{} - \sum_{q} \Lambda \d (q) \mathcal{D} (q)^{-1}_{} \Lambda (q)
	\en
	where $\tilde{\action} _{}^{SP} = \beta F_{}^{SP}$, $\Lambda (q)$ is a column vector composed of different fluctuation fields, and $\mathcal{D}(q)$ is the fluctuation propagator matrix. For small amplitude and phase fluctuations, $\eta _\alpha ^{} (q) \simeq (-1)^\alpha_{} \Delta_{SP}^{} \left(\zeta _\alpha ^{} (q) + i\theta _\alpha ^{} (q) \right)$, where $\zeta _\alpha ^{} (q)$ and $\theta _\alpha ^{} (q)$ are amplitude and phase fluctuation fields, respectively. On changing the basis to symmetric and anti-symmetric modes (e.g.,~via $\zeta_{x,y}(q) = \zeta_s(q) \pm \zeta_a(q)$ and $\theta_{x,y}(q) = \theta_s(q) \pm \theta_a(q)$, see \cite{MandalPRB00}),  $\mathcal{D}(q=0)^{-1}_{}$ becomes diagonal. The gap parameters of the fluctuation modes, $\MSAM$, $\MAPM$ and $\MAAM$ are the appropriate diagonal entries of $\mathcal{D}(q=0)^{-1}_{}$. Further, an inspection of $\mathcal{D}(q)^{-1}_{}$ for small values of $q$ gives $\rho_{s}^{}$ (see SM  S1). It is to be noted that our approach leads to a ``free energy functional'' consistent with the Ginzburg-Landau approach of \cite{BerlinskyPRB94,BerlinskyPRL95} (see SM S4).

\begin{figure}
\centering
\includegraphics[width=1.0\linewidth]{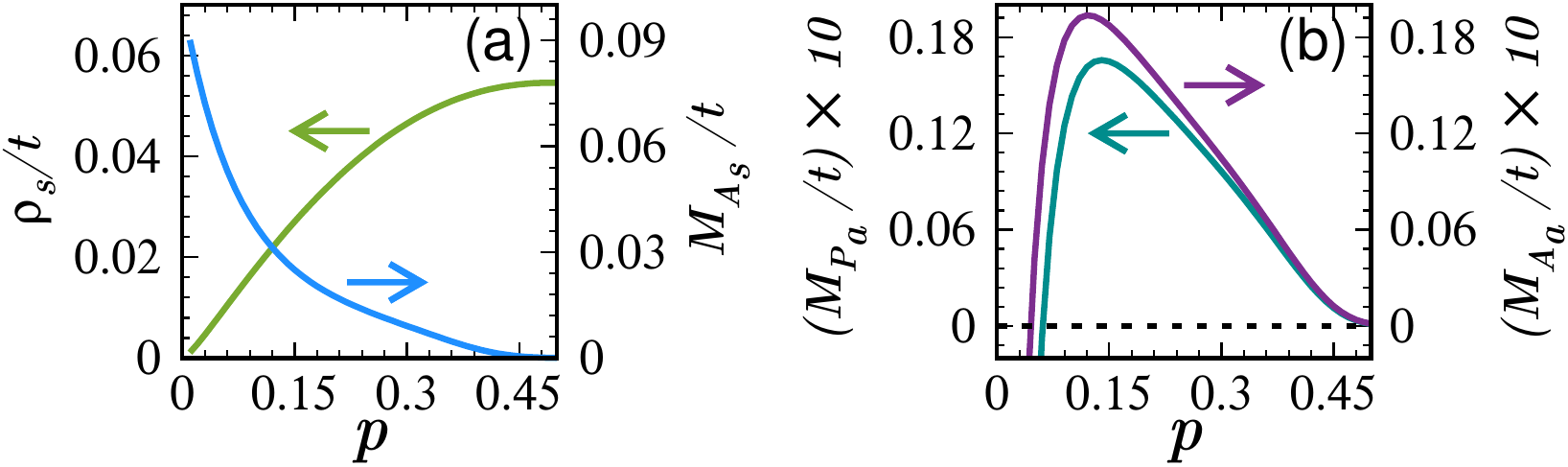}
\caption{(color online) \textbf{Properties of collective modes (analytic):} Doping dependence of {\bf (a)} $\rho_s^{}/t$, the superfluid stiffness of the $\SPM$, and $\MSAM/t$, the gap parameter of the $\SAM$, {\bf (b)} $\MAPM/t$, the gap parameter of the $\APM$, and $\MAAM/t$, the gap parameter of the $\AAM$.}
\label{fig:proj_scales}
\end{figure}

\Fig{fig:proj_scales} shows the evolution of the nature of the collective excitations as a function of hole doping $\doping$. \Fig{fig:proj_scales}(a) reproduces well known results -- the superfluid stiffness $\rho_s$ increases linearly with $\doping$, and the gap parameter associated with the $\SAM$ increases up on decreasing $\doping$. \Fig{fig:proj_scales}(b) shows one of the key findings of this paper. The $\APM$ gap parameter $\MAPM$ increases up on decreasing $\doping$, attaining a maximum around $\doping \approx 0.14$ and then has a precipitous fall. In fact, {\em $\MAPM$ goes to zero at a finite critical doping $\doping_c \approx 0.06$}! Interestingly, the gap parameter associated with the $\AAM$ follows suit, closing at a doping slightly less than $\doping_c$ (\Fig{fig:proj_scales}(b)). These results clearly indicate that the $d$-SC state is not stable at lower values of $\doping$ owing to the fluctuations of the internal anti-symmetric modes, i.e., the $\APM$ and the $\AAM$.

The physics of this can be traced to the {\em effect of strong correlations} encoded in the Gutzwiller factors. For example, at $T = 0$
	\be
	\label{eq:MAPM}
	\MAPM = 4\Delta^2_{}\left[ \frac{1}{J_P^{}} - \frac{1}{2\sites} \sum_{\veck} \frac{(\ckxpcky) ^2_{}}{E_\veck^{}} \right]
	\en
	While the first term in RHS is positive and is proportional to $1/J_P^{}(\doping)$, the second term is negative and is proportional to $1/g_t^{}(\doping) t$. As doping $p \rightarrow 0$, $J_P^{}(\doping) \sim J$ and $g_t^{}(\doping) t \sim 0$, making $\MAPM$ negative. A similar analysis explains the softness of $\AAM$. In contrast, the $\SAM$ gap parameter is manifestly positive (see SM S1). Moreover, these results do not depend on the choice of Gutzwiller factors as shown in SM S1.

\noindent
{\bf VMC Investigation:} This remarkable result motivated us to estimate the gap parameter of the anti-symmetric modes using variational Monte Carlo (VMC) method\cite{GrosAdvPhys07}.  We study  the $t$-$J$ model
{\small	\be
	\label{eq:H_tJ}
	H_{tJ}^{} = - \sum_{\stackrel{i, \boldsymbol \delta}{\sigma}} t(\boldsymbol \delta) \mathcal{P} c_{i+\boldsymbol \delta \sigma} \d c_{i \sigma}^{} \mathcal{P} + J\sum_{\langle i,j \rangle} \left( \mathbf{S}_i \cdot \mathbf{S}_j - \frac{1}{4} n_i n_j \right)
	\en}
	where the projection operator $\mathcal{P}$ restricts hopping processes from exploring doubly occupied states\cite{FukuyamaRPP08}. All other symbols (parameters) have the same meanings (values) as in the section above. The ``ground state'' of this system with $d$-SC pairing can be obtained by constructing an appropriate ``projected BCS state'' (see, e.~g., \cite{PathakPRL09}) described by two variational parameters $\Delta^{v}$ and $\mu ^{v}$. 

	An estimation of the parameters associated with the collective excitations requires the study of {\em excited} states. Estimation of $\rho_s$ using such approaches has proved to be challenging\cite{AltmanPRL10}. Our interest here is in the {\em anti-symmetric} modes that are homogeneous over the unit cells (i.~e., $\mathbf{q} = \bzero$ in the notation of the previous section).  To study these excitation, we develop an ansatz for the excited states corresponding to the anti-symmetric modes. We take the pairing variational parameters on the $x$-bonds to be of the form $\Delta^v_x = \Delta^v_0 \left( 1 + \zeta \right)e^{+i \theta}$, and $ \Delta^v_y = -\Delta^v_0 \left( 1 - \zeta \right)e^{-i \theta}$ on the $y$-bonds, where $\Delta_0^v$ is the parameter corresponding to variational ground state. This leads to a translationally invariant excited state wavefunction that is computationally tractable. The gap parameter corresponding to the anti-symmetric modes can be {\em estimated} by computing the second derivatives of the energy of this state with respect to the parameters $\theta$ and $\zeta$. For example, $\left. \MAPM \right|_{\mbox{\tiny VMC}} = \left. \frac{\dou^2 E_{VMC}(\zeta=0,\theta)}{\dou \theta^2} \right|_{\theta = 0}$. It is very important to keep in mind that these quantities provide an {\em upper bound} for the gap parameters associated with these modes, as the variational ansatz does not include all the quantum fluctuations\footnote{There are also significant numerical challenges in estimating the gap parameters, since large deviation from the d-SC state is involved}. 

\begin{figure}
\centering
\includegraphics[width=0.9\linewidth]{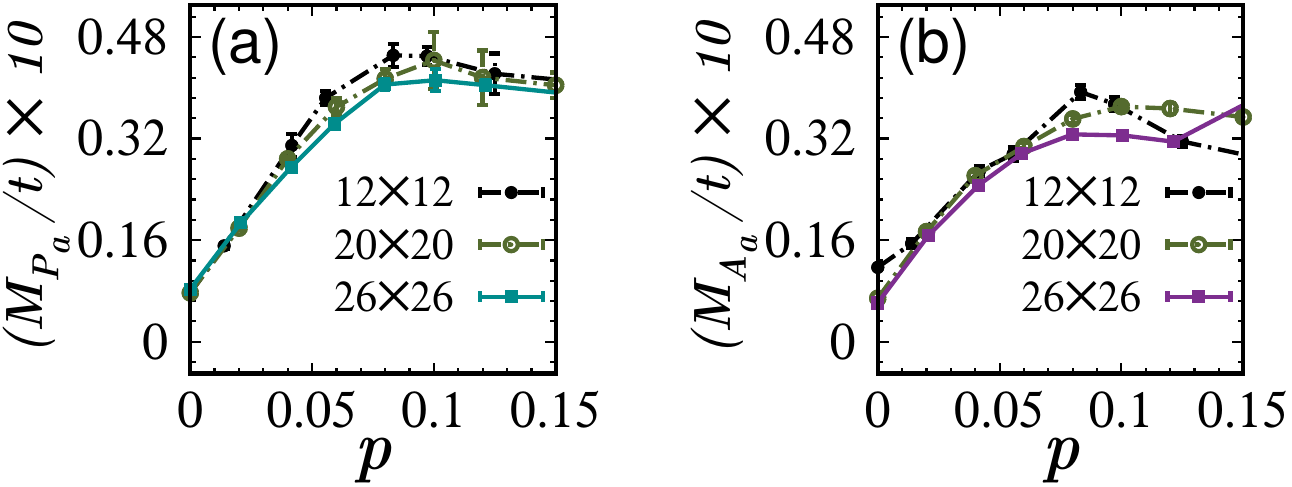}
\caption{(color online) \textbf{Gap parameters obtained from VMC:} Doping dependence of the upper bounds on {\bf (a)} $\MAPM/t$ and {\bf (b)} $\MAAM/t$.}
\label{fig:VMC_SBMFT}
\end{figure}

\Fig{fig:VMC_SBMFT} shows the estimates for the upper bounds of gap parameters of the anti symmetric modes obtained from the VMC calculations -- reassuringly, the finite size effects are minimal. Most interestingly, not only are the orders of magnitudes of the gap parameters similar to that found in the previous section, they also have a similar qualitative non-monotonic behavior as a function of the hole doping $\doping$. Just as in the previous section, there is a precipitous fall in the gap parameters with decreasing hole doping for $p \lesssim 0.1$. Indeed, the results of the previous section are consistent with the bounds provided by the VMC calculations.

\noindent
{\bf Phase Diagram -- Large-$\mathcal{N}$ formalism:} How do the instabilities uncovered above revise the phase diagram? We address this question using a large-$\mathcal{N}$ formulation\cite{RadzihovskyPRA07}. Such formalisms have been used before to identify the different phases in $t$-$J$ like models, and to study the transitions between these phases\cite{SachdevPRL99,SachdevPRB00}. In contrast, here, our main focus is to study how the saddle point $d$-wave pairing amplitude is itself renormalized by the quantum fluctuations discussed above. For this, we study a $\mathcal{N}$ flavor generalization of the Hamiltonian defined in \eqn{eq:H_Gutz}.
\begin{multline}
	\label{eq:H_large-N}
	H = -g_t^{}(p) \sum_{\stackrel{i, \boldsymbol \delta, \sigma}{\lambda=1}}^{\lambda=\mathcal{N}} t(\boldsymbol \delta) c_{i+\boldsymbol \delta \sigma \lambda} \d c_{i \sigma \lambda}^{} -\frac{J_P^{}}{\mathcal{N}}\sum_{\langle i,j \rangle} b_{ij}^{\mathcal{N} \dagger} b_{ij}^{\mathcal{N}}\\
	-\frac{J_K^{}}{\mathcal{N}}\sum_{\langle i,j \rangle} \chi_{ij}^{\mathcal{N} \dagger} \chi_{ij}^{\mathcal{N}} - \mu \sum_{\stackrel{i, \sigma}{\lambda=1}}^{\lambda=\mathcal{N}} c_{i \sigma \lambda} \d c_{i \sigma \lambda}^{}
	\end{multline}
	where the fermions now carry an additional flavor index $\lambda \in \{1,2,\ldots,\flavours \}$. $J_P^{} = J(g_s^{}+g_n^{})/2$, and $b_{ij}^{\mathcal{N}\dagger} = \sum _{\lambda=1}^\mathcal{N} (c_{i \uparrow \lambda} \d c_{j \downarrow \lambda} \d - c_{i \downarrow \lambda} \d c_{j \uparrow \lambda} \d)/\sqrt{2}$ is the pair creation operator. While, $J_K^{} = J(g_s^{}-g_n^{})/2$, and $\chi_{ij}^{\mathcal{N}\dagger} = \sum _{\lambda=1}^\mathcal{N} (c_{i \uparrow \lambda}\d c_{j \uparrow \lambda}^{} + c_{i \downarrow \lambda}\d c_{j \downarrow \lambda}^{})/\sqrt{2}$ is the ``Fock'' operator. $p$ is the hole doping of each of the $\lambda$ flavor fermions. All other symbols (parameters) have the same meanings (values) as before. 

\begin{figure}
\centering
\includegraphics[width=0.7\linewidth]{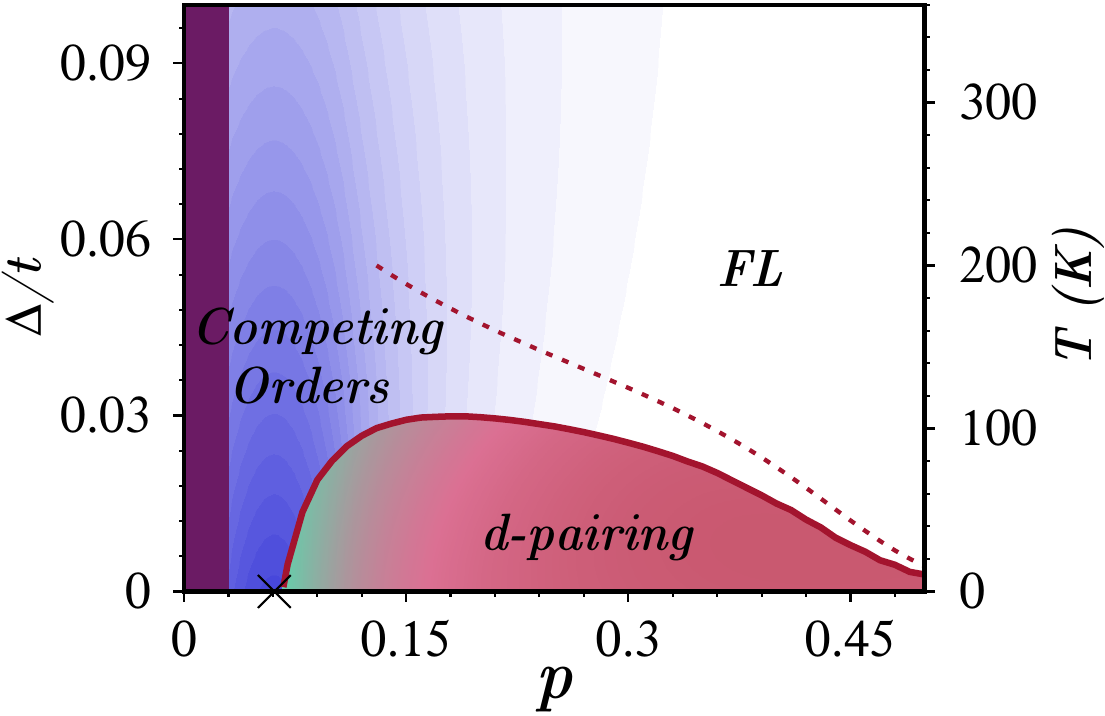}
\caption{(color online) \textbf{Phase diagram ($1/\flavours$ theory):} $d$-pairing scale ($\Delta/t$) from $1/\flavours$ theory (solid red curve), and its saddle point value (dashed red line). Cross on doping axis marks the $\APM$ instability. Using $t = 3600K$, inferred finite $T$ phase diagram is also depicted. Fermi liquid ({\em FL}) behavior is expected at large $p$ and $T$ above the $d$-pairing scale. At small $p$, $d$-pairing is unstable to $\APM$ fluctuations and ``normal state'' hosts {\em competing orders} (blue shade). Light green region depicts fluctuation dominated $d$-SC. And, purple region has anti-ferromagnetism.}
\label{fig:large-N}
\end{figure}

A detailed analysis (see SM S3) gives $1/\mathcal{N}$ corrections, $\delta \Delta$, $\delta K$ and $\delta \mu$, to the saddle point ($\flavours \rightarrow \infty$) values of $\Delta$, $K$ and $\mu$, respectively. Note that the renormalized $\Delta$ obtained in our calculations is the fluctuation corrected $d$-pairing scale. \Fig{fig:large-N} shows the results obtained from this large-$\flavours$ theory ($\Delta /t$ \textit{vs} $p$). On the overdoped side $p > 0.2$, where $\rho_s^{}$ is large, fluctuations play only a quantitative role. On the other hand the picture is drastically changed on the underdoped side. The internal anti-symmetric fluctuations drastically reduce the $d$-pairing scale forcing it to vanish around $p_c^{} \approx 0.06$. That is, with the inclusion of the internal modes of fluctuations, $d$-SC ceases to be the ground state for $p < p_c^{}$. It must be appreciated that this is so apart from considerations of a competing AF order, due to the nature of excitations of the $d$-SC state itself\footnote{Such a possibility is touched up on by G.~Kotliar in \cite{Kotliar1PRB88}, but is not explored in depth.}.

These results also allow us to infer the finite $T$ phase diagram of cuprates (\Fig{fig:large-N}). At large doping, the ground state is a $d$-SC with small pairing strength, and the normal state here is expected to have a Fermi liquid like character. At intermediate dopings, where $\rho_s^{}$ becomes comparable to the fluctuation corrected $\Delta$, the physics of phase incoherent ``preformed $d$-pairs'' is likely to be important below the $d$-pair breaking temperature. Whereas, at low dopings the $d$-SC phase is destroyed by the weakening of the gaps of the internal modes. Our theory suggests that the finite $T$ phase near $p=p_c^{}$ will have large ``critical'' fluctuations corresponding to mixing in extended $s$ like order. Needless to say, interactions (e.g. long range coulomb interaction) can become operative in this regime, possibly stabilizing other orders such as CDW\cite{SachdevPRL99}. This provides a crucial insight into the origin of many types of orders seen in the underdoped side. Note that this scenario for cuprates is similar to what happens in heavy fermion systems where superconductivity (competing order) ``covers up'' the quantum critical points\cite{NormanScience11}. Furthermore, since the renormalized $d$-pair breaking scale found here is much smaller than the experimental pseudogap scale, we infer that the pseudogap regime will not only be influenced by the $d$-pairing, but also by other competing orders, including non-$d$ pairing.

\noindent
{\bf Discussion:} STM studies of cuprates have established the existence of two gap scales in underdoped samples\cite{KimNatLett10}. Recently, Hoffman \textit{et al}\cite{HoffmanScience14} have been able to distinguish the $d$-SC gap from the pseudogap, for overdoped Bi2201. This indicates that $d$-pairing alone may not be sufficient to cause the pseudogap. This view is bolstered by our $1/\flavours$ result for the fluctuation corrected $d$-pairing scale $\Delta (p) /t $ (\Fig{fig:large-N}), which closes at $p=p_c^{}$.

Vishik \textit{et al}\cite{ShenPNAS12}, in an ARPES study of Bi2212, report the doping dependence of the near nodal gap (see Fig.~2(d) of \cite{ShenPNAS12}) to have a form very similar to the fluctuation corrected $\Delta (p)/t$ in \Fig{fig:large-N}. In particular, they find a ``saturation'' of the near nodal gap with decreasing doping. This ``saturation'' is qualitatively observed in \Fig{fig:large-N}, and, in our study, owes to the small gap parameters associated with the internal modes. The QCP in \cite{ShenPNAS12} close to $p\sim0.07$ may be related to the $\APM$ instability that we have uncovered, and the fully gapped phase observed for smaller doping to the $d+is$ order that the instability points towards (see SM S4). Such a correspondence with experiment is encouraging, and we hope to explore it further in future work.

It will be interesting also to explore the possibilities of observing the anti-symmetric fluctuation modes directly in an experiment. We show in SM S4, using an appropriate Ginzburg-Landau functional\cite{BerlinskyPRB94,BerlinskyPRL95}, that the presence of charge currents modifies the gap parameter of the $\APM$ ($\MAPM$). This implies that the associated $p_c^{}$ would also be sensitive to charge currents in the copper oxide plane. Furthermore, for $p>p_c^{}$ the availability of low lying amplitude fluctuations should be detectable in Tera-hertz spectroscopy experiments\cite{MatsunagaPRL13,ShermanNatPhys15}. Low temperature electronic specific heat measurements should also bear signatures of these internal anti-symmetric fluctuation modes.

In conclusion, the main message of this paper is that strong correlation induced softness of the anti-symmetric phase and amplitude fluctuations of the $d$-SC order parameter in cuprate superconductors make the $d$-SC state intrinsically fragile in the underdoped side. From the result that the $d$-pairing gap is driven to zero at $p_c^{}$, we infer that the experimentally observed pseudogap has contributions from mechanisms other than $d$-pairing. As discussed, this point not only throws light on known experimental results, but also suggests new ones towards uncovering the cuprate enigma. From a theoretical perspective, this study suggests further new directions. It would be interesting to develop and study the critical theory of the QCP uncovered here. Also, how the presence of nodal quasi-particles affect the physics described above needs to be explored. Away from the QCP, developing approaches to study the excited states, including order parameter fluctuations that are not treated explicitly in VMC, would be important.

	{\bf Acknowledgements:} The authors thank T.~V.~Ramakrishnan and Sri Raghu for discussions, Masao Ogata and Arun Paramekanti for communications. AVM thanks Pinaki Majumdar, Arijit Haldar and Sumilan Banerjee for stimulating interactions. This research has been supported by several funding agencies of the Government of India (AVM(CSIR--SPMF), UKY(UGC--DSKPDF), AM(DST), HRK(DST), VBS(DST/DAE)) the authors thank all of them for the support.

\bibliography{bib_apm}	


\clearpage
\newpage

\appendix

\renewcommand{\appendixname}{}	
\renewcommand{\thesection}{{S\arabic{section}}}
\renewcommand{\thefigure}{\thesection.\arabic{figure}}
\renewcommand{\theequation}{\thesection.\arabic{equation}}

\setcounter{page}{1}
\setcounter{equation}{0}
\setcounter{figure}{0}

\widetext

\centerline{\bf SUPPLEMENTAL MATERIAL}
\centerline{ }
\centerline{\bf for}
\centerline{ }
\centerline{\bf \ourtitle}
\centerline{ }
\centerline{Aabhaas V. Mallik$^1$, Umesh K. Yadav$^1$, Amal Medhi$^2$, H.~R. Krishnamurthy$^1$, and Vijay B. Shenoy$^1$}
\centerline{ }
\centerline{\it $^1$Department of Physics, Center for Condensed Matter Theory, Indian Institute of Science, Bengaluru - 560012, India}
\centerline{ }
\centerline{\it $^2$School of Physics, Indian Institute of Science Education and Research, Thiruvananthapuram - 695016, India}

	\section{Fluctuation scales}
	\label{Appen:fluc_scales}

	In this section we describe the computation of the gap parameters of the fluctuation modes and the superfluid density starting from \eqn{eq:H_Gutz}. The action, corresponding to the Matsubara-momentum space version of \eqn{eq:H_Gutz}, can be written in terms of the Grassmann fields ($\bar{c}_{k \sigma}^{}$ and $c_{k \sigma}^{}$) as
	\begin{equation}
		\label{eq:S1_action}
		\action \left[ \{ \bar{c}_{k' \sigma'}^{}, c_{k' \sigma'}^{} \}\right]= \sum _{k,\sigma} \bar{c}_{k \sigma}^{} \left( -ik_n + \xi_\veck^{} \right) c_{k \sigma}^{}
		- \frac{J_P^{}}{\sites \beta} \sum _{q,\alpha} \bar{b}_\alpha^{}(q) b_\alpha^{}(q) - \frac{J_K^{}}{\sites \beta} \sum _{q,\alpha} \bar{\chi}_\alpha^{}(q) \chi_\alpha^{}(q)
	\end{equation}
where $\xi_\veck^{} = -\sum_{\boldsymbol \delta} g_t^{} t_{\boldsymbol \delta}^{} \exp(i\veck \cdot \boldsymbol \delta) - \mu $, $\beta = 1/T$, $\mu$ is the chemical potential,  and $N$ is the number of lattice sites; $k = \left( ik_n^{}, \veck \right)$ and $q = \left( iq_l^{}, \vecq \right)$, with $ik_n^{}$ and $iq_l^{}$, respectively, being fermionic and bosonic Matsubara frequencies. $J_P^{} = J(g_s^{}+g_n^{})/2$, $\bar{b}_\alpha ^{} (q) = \sum_\veck \bar{c}_{k+q \uparrow}^{} \bar{c}_{-k \downarrow}^{} a^*_\alpha (\vecq, \veck)$ is the pair creation field with $\alpha \in \{ 0,1 \}$ or $\{ x,y \}$ as appropriate, and $a^{}_\alpha\left( \vecq , \veck \right) = \sqrt{2}\cos\left( k_\alpha ^{} + q_\alpha^{}/2\right) \exp\left( iq_\alpha^{}/2\right)$. Similarly, $J_K^{} = J(g_s^{}-g_n^{})/2$ and $\bar{\chi}_\alpha ^{} (q) = \sum_\veck \exp(i\veck \cdot \mathbf{e_\alpha^{}}) \left( \bar{c}_{k+q \uparrow}^{} c_{k \uparrow}^{} + \bar{c}_{k+q \downarrow}^{} c_{k \downarrow}^{} \right)/\sqrt{2}$ is ``Fock'' field.

	By introducing the Hubbard-Stratonovich fields $\Delta_{\alpha}^{}(q)$ and $K_{\alpha}^{}(q)$ we make the action quadratic in Grassmann fields ($\bar{c}_{k \sigma}^{}$, $c_{k \sigma}^{}$)
	\begin{equation}
		\label{eq:S1_action_quad}
		\action \left[ \{ \Psi_{k' \sigma'}^{}, \Delta_{\alpha'}^{}(q'), K_{\alpha'}^{}(q') \} \right]= \frac{1}{J_P^{}} \sum _{q,\alpha} |\Delta_\alpha ^{} (q) |^2_{} + \frac{1}{J_K^{}} \sum _{q,\alpha} |K_\alpha ^{} (q) |^2_{} - \sum_{k,k'} \bar{\Psi}_k^{} G^{-1}_{}(k,k') \Psi_{k'}^{}
	\end{equation}
	with
	\begin{equation}
	\bar{\Psi}_k^{} = \left( \bar{c}_{k\uparrow}^{}\qquad c_{-k\downarrow}^{}\right)
	\end{equation}
	and
	\begin{equation}
		\label{eq:S1_Gkkp}
		-G_{}^{-1} (k,k') =
		\begin{pmatrix}
		\begin{split} & \left(-ik_n^{} + \xi_\veck^{}\right) \delta _{kk'}^{} \\
		&-\frac{1}{\sqrt{2 \sites\beta}}\sum_\alpha \left( \text{e}^{i \veck ' . \mathbf{e}_\alpha^{}}_{} K_\alpha^{} (k-k') + \text{e}^{- i \veck  . \mathbf{e}_\alpha^{}}_{} K_\alpha^{*} (k'-k) \right)\end{split} \qquad
		-\frac{1}{\sqrt{\sites\beta}}\sum_\alpha a^*_\alpha(\veck-\veckp,\veckp)\Delta_\alpha^{}(k-k')\\
		\\
		-\frac{1}{\sqrt{\sites\beta}}\sum_\alpha a^{}_\alpha(\veckp-\veck,\veck)\Delta_\alpha^{*}(k'-k) \qquad
		\begin{split} & -\left(ik_n^{} + \xi_\veck^{}\right) \delta _{kk'}^{} \\
		&-\frac{1}{\sqrt{2 \sites\beta}}\sum_\alpha \left( \text{e}^{-i \veck . \mathbf{e}_\alpha^{}}_{} K_\alpha^{} (k-k') + \text{e}^{i \veck '. \mathbf{e}_\alpha^{}}_{} K_\alpha^{*} (k'-k) \right)\end{split}
		\end{pmatrix}
	\end{equation}
	On integrating out the quadratic Grassmann fields we obtain
	\begin{equation}
	\label{eq:S1_action_bosonic}
	\tilde{\action} \left[ \{ \Delta_{\alpha'}^{}(q'), K_{\alpha'}^{}(q') \} \right]= \frac{1}{J_P^{}} \sum _{q,\alpha} |\Delta_\alpha ^{} (q) |^2_{} + \frac{1}{J_K^{}} \sum _{q,\alpha} |K_\alpha ^{} (q) |^2_{} - \ln \left[ \det \left(-G^{-1}_{} \right) \right]
	\end{equation}
	The saddle point equations for fields $\Delta_\alpha^{}(q)$ and $K_\alpha^{}(q)$ are then obtained to be
	\begin{equation}
	\label{eq:S1_SPeq_Delta}
	\frac{\delta \tilde{\action}}{\delta \Delta_\alpha^{*}(q)} = \frac{\Delta_\alpha^{}(q)}{J_P^{}} -\frac{1}{\sqrt{\sites \beta}} \sum_k a_\alpha^{}(\vecq ,\veck) G_{12}^{}(k+q,k) = 0
	\end{equation}
	and
	\begin{equation}
	\label{eq:S1_SPeq_K}
	\frac{\delta \tilde{\action}}{\delta K_\alpha^{*}(q)} = \frac{K_\alpha^{}(q)}{J_K^{}} -\frac{1}{\sqrt{2 \sites \beta}} \sum_k \left[ \text{e}^{-i \veck . \mathbf{e}_\alpha^{}}G_{11}^{}(k+q,k) - \text{e}^{i (\veck + \vecq) . \mathbf{e}_\alpha^{}}G_{22}^{}(k+q,k) \right] = 0
	\end{equation}
	To solve for an uniform $d$-wave pairing saddle point we set $\Delta_\alpha^{}(q) = \sqrt{\sites \beta} (-1)^\alpha_{} \Delta_{SP} \delta_{q,0}$ and $K_\alpha^{}(q) = \sqrt{\sites \beta} K_{SP} \delta_{q,0}$ in \eqn{eq:S1_SPeq_Delta} and \eqn{eq:S1_SPeq_K}, which gives
	\begin{equation}
	\label{eq:S1_SP_Delta}
	\Delta_{SP}^{} = \frac{J_P^{}}{4\sites} \sum_\veck \sqrt{2}(\cos{k_x^{}} - \cos{k_y^{}}) \frac{\Delta_\veck^{}}{E_\veck^{}} \tanh{\frac{\beta E_\veck^{}}{2}}
	\end{equation}
	and
	\begin{equation}
	\label{eq:S1_SP_K}
	K_{SP}^{} = -\frac{J_K^{}}{4\sites} \sum_\veck \sqrt{2}(\cos{k_x^{}} + \cos{k_y^{}}) \frac{\tilde{\xi}_\veck^{}}{E_\veck^{}} \tanh{\frac{\beta E_\veck^{}}{2}}
	\end{equation}
	where
	\begin{equation}
	\label{eq:S1_E_k}
	E_\veck^{} = \sqrt{\tilde{\xi}_\veck^{2} + \Delta_{\veck}^{2}}
	\end{equation}
	\begin{equation}
	\label{eq:S1_xi_k}
	\tilde{\xi}_\veck^{} = \xi_\veck^{} - K_\veck^{}
	\end{equation}
	\begin{equation}
	\label{eq:S1_K_k}
	K_\veck^{} = \sqrt{2}K_{SP}^{}(\cos{k_x^{}} +\cos{k_y^{}})
	\end{equation}
	and
	\begin{equation}
	\label{eq:S1_Delta_k}
	\Delta_\veck^{} = \sqrt{2}\Delta_{SP}^{}(\cos{k_x^{}} -\cos{k_y^{}})
	\end{equation}

	Next, to study the fluctuations in the $d$-pairing field we set  $\Delta_\alpha^{}(q) = \sqrt{\sites \beta} (-1)^\alpha_{} \Delta_{SP} \delta_{q,0} + \eta_\alpha^{}(q)$ in \eqn{eq:S1_action_bosonic} and expand up to quadratic order in $\eta$ to obtain,
	\begin{equation}
	\label{eq:S1_action_gaussian}
	\tilde{\action} \left[ \{ \eta _{\alpha'} ^{*} (q'), \eta _{\alpha'} ^{} (q')\} \right] \simeq \tilde{\action} ^{SP}_{} - \sum_{q} \Lambda \d (q) \mathcal{D} (q)^{-1}_{} \Lambda (q)
	\end{equation}
	where
	\begin{equation}
	\label{eq:S1_S_SP}
	\tilde{\action} ^{SP}_{} = \frac{2 \sites \beta}{J_P^{}} \Delta_{SP}^2 + \frac{2 \sites \beta}{J_K^{}} K_{SP}^2 + \beta \sum_{\bk} \left( \tilde{\xi} _{\bk}^{} - E_{\bk}^{} \right) - 2 \sum_{\bk} \ln \left( 1 + \text{e}^{-\beta E_{\bk}^{}} \right) = \beta F^{SP}_{}
	\end{equation}
	\begin{equation}
	\label{eq:S1_fluc_vector}
	\Lambda \d (q) = \left( \eta _{0} ^{*} (q) \quad \eta _{0} ^{} (-q) \quad \eta _{1} ^{*} (q) \quad \eta _{1} ^{} (-q) \right)
	\end{equation}
	and
	\begin{equation}
	\label{eq:S1_fluc_prop}
	\mathcal{D} (q)^{-1}_{} = -
	\begin{pmatrix}
		\frac{1}{2J_P^{}} + C_{00}^{}(q) & B_{00}^{}(q) & C_{01}^{}(q) & B_{01}^{}(q)\\ \\
		B_{00}^{}(q) & \frac{1}{2J_P^{}} + C_{00}^{}(-q) & B_{01}^{}(q) & C_{10}^{}(-q)\\ \\
		C_{10}^{}(q) & B_{10}^{}(q) & \frac{1}{2J_P^{}} + C_{11}^{}(q) & B_{11}^{}(q)\\ \\
		B_{10}^{}(q) & C_{01}^{}(-q) & B_{11}^{}(q) & \frac{1}{2J_P^{}} + C_{11}^{}(-q)
	\end{pmatrix}
	\end{equation}
	is the inverse fluctuation propagator with
	\begin{equation}
	\label{eq:S1_DC}
	C_{\alpha \gamma}^{}(q) = \frac{1}{2\sites \beta} \sum_k a_\gamma^{*}(\vecq , \veck) a_\alpha^{}(\vecq , \veck) G_{11}^{SP}(k+q,k+q) G_{22}^{SP}(k,k)
	\end{equation}
	\begin{equation}
	\label{eq:S1_DB}
	B_{\alpha \gamma}^{}(q) = \frac{1}{2\sites \beta} \sum_k a_\gamma^{*}(\vecq , \veck) a_\alpha^{}(\vecq , \veck) G_{12}^{SP}(k+q,k+q) G_{12}^{SP}(k,k)
	\end{equation}
	and
	\begin{equation}
		\label{eq:S1_GSP}
		G_{}^{SP} (k,k') =
		\begin{pmatrix}
		ik_n^{} + \tilde{\xi}_\veck^{} & -\Delta_{\veck}^{}\\
		\\
		-\Delta_{\veck}^{} & ik_n^{} - \tilde{\xi}_\veck^{} \\
		\end{pmatrix}
		\frac{\delta _{kk'}^{}}{(ik_n^{})^2_{} - E_\veck^2}
	\end{equation}
	as obtained by setting $\Delta_\alpha ^{} (q) = (-1)^\alpha_{} \sqrt{\sites \beta }\Delta_{SP}^{} \delta _{q,0}^{}$ and $K_\alpha ^{}(q) = \sqrt{\sites \beta }K_{SP}^{} \delta _{q,0}^{}$ in \eqn{eq:S1_Gkkp}.

\begin{figure}
\centering
\includegraphics[width=1.0\linewidth]{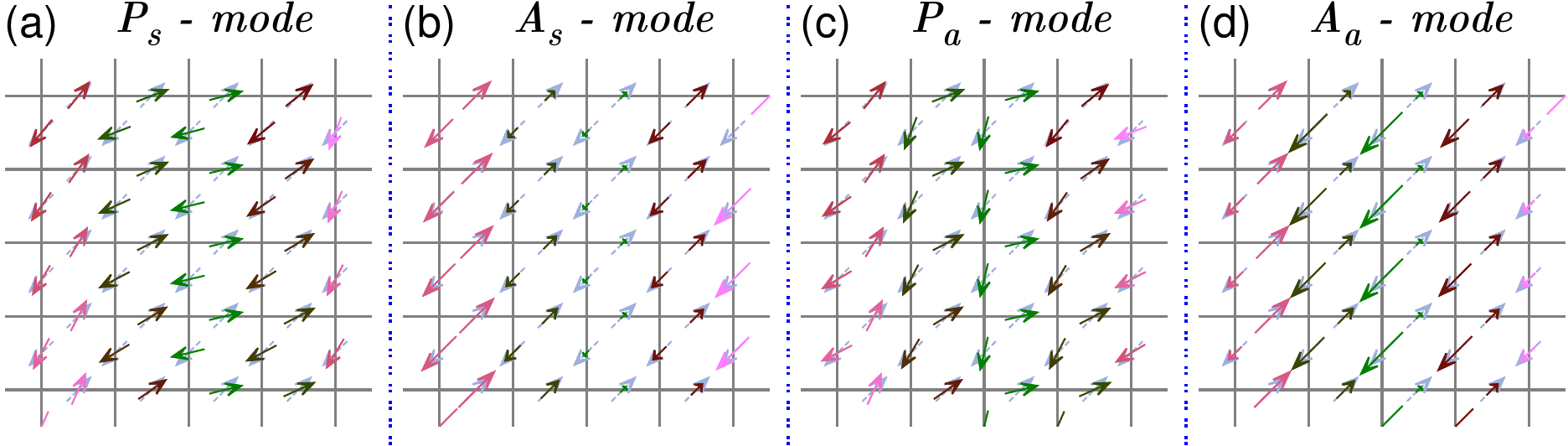}
\caption{(color online) \textbf{Collective modes of $\boldsymbol{d}$-SC:} Length and direction of an arrow, respectively, depict the magnitude and phase of the local pairing order. Dashed grey arrows in the background show the uniform d-wave order. Different fluctuation modes about this state with wave vector $\mathbf{q}=\left(\pi/3,\pi/24 \right)$ are shown in {\bf (a)} Symmetric phase ($\SP$) mode, {\bf (b)} Symmetric amplitude ($\SA$) mode, {\bf (c)} Anti-symmetric phase ($\AP$) mode, and {\bf (d)} Anti-symmetric amplitude ($\AA$) mode.}
\label{fig:schematic}
\end{figure}

	When $\mathcal{D}(q)$ is diagonalized, its poles give the dispersion of the collective modes of the system. This is straight forward to do when $q = 0$ with small amplitude and phase fluctuations ($\eta _\alpha ^{} (q) \simeq (-1)^\alpha_{} \Delta_{SP}^{} \left(\zeta _\alpha ^{} (q) + i\theta _\alpha ^{} (q) \right)$, where $\zeta _\alpha ^{} (q)$ and $\theta _\alpha ^{} (q)$ are amplitude and phase fluctuation fields, respectively). In terms of the symmetric and anti-symmetric modes (e.g.~via $\zeta_{x,y}(q) = \zeta_s(q) \pm \zeta_a(q)$ and $\theta_{x,y}(q) = \theta_s(q) \pm \theta_a(q)$, see \cite{MandalPRB00} and \Fig{fig:schematic}) we find,

	\begin{equation}
	\label{eq:S1_D_0}
	- \Lambda \d (q=0) \mathcal{D} (0)^{-1}_{} \Lambda (0) = - \tilde{\Lambda}\d (0) \tilde{\mathcal{D}} (0)^{-1}_{} \tilde{\Lambda}(0)
	\end{equation}
	where
	\begin{equation}
	\label{eq:S1_fluc_vector_0}
		\tilde{\Lambda}\d (0) = \left( \zeta _{a} ^{*} (0) \quad \zeta _{s} ^{*} (0) \quad \theta _{a} ^{*} (0) \quad \theta _{s} ^{*} (0) \right)
	\end{equation}
	and
	\begin{equation}
	\label{eq:S1_fluc_prop_0}
	- \tilde{\mathcal{D}} (q=0)^{-1}_{} = \frac{1}{2}
	\begin{pmatrix}
		\MAAM & 0 & 0 & 0\\
		0 & \MSAM & 0 & 0\\
		0 & 0 & \MAPM & 0\\
		0 & 0 & 0 & 0
	\end{pmatrix}
	\end{equation}

	Clearly, the $\SPM$ is the gapless (massless) Goldstone mode arising due the breaking of the continuous $U(1)$ symmetry. Whereas, the $A_{a}$, $A_{s}$ and $P_{a}$ modes are gapped (massive) with gap parameters $\MAAM$, $\MSAM$ and $\MAPM$, respectively. We find
	\begin{subequations}
	\begin{align}
		\label{eq:S1_MAPM}
		\MAPM (T = 0) &= 4\Delta^2_{}\left[ \frac{1}{J_P^{}} - \frac{1}{2\sites} \sum_{\veck} \frac{(\ckxpcky) ^2_{}}{E_\veck^{}} \right] \\
		&= -\frac{8\Delta^2_{}}{\sites} \sum_{\veck} \frac{ \ckxcky }{E_\veck^{}}
	\end{align}
	\end{subequations}
	\begin{equation}
		\label{eq:S1_MAAM}
		\MAAM (T = 0) = \MAPM (T = 0) + \frac{4\Delta^4_{}}{\sites}\sum_{\veck} \frac{\left( \cos^2 k_x^{} - \cos^2 k_y^{}\right)^2}{E_\veck^3}
	\end{equation}
	\begin{equation}
		\label{eq:S1_MSAM}
		\MSAM (T = 0) = \frac{4\Delta^4_{}}{\sites}\sum_{\veck} \frac{\left( \ckxmcky \right)^4}{E_\veck^3}
	\end{equation}
	\begin{figure}
		\centering
		\includegraphics[width=1.0\linewidth]{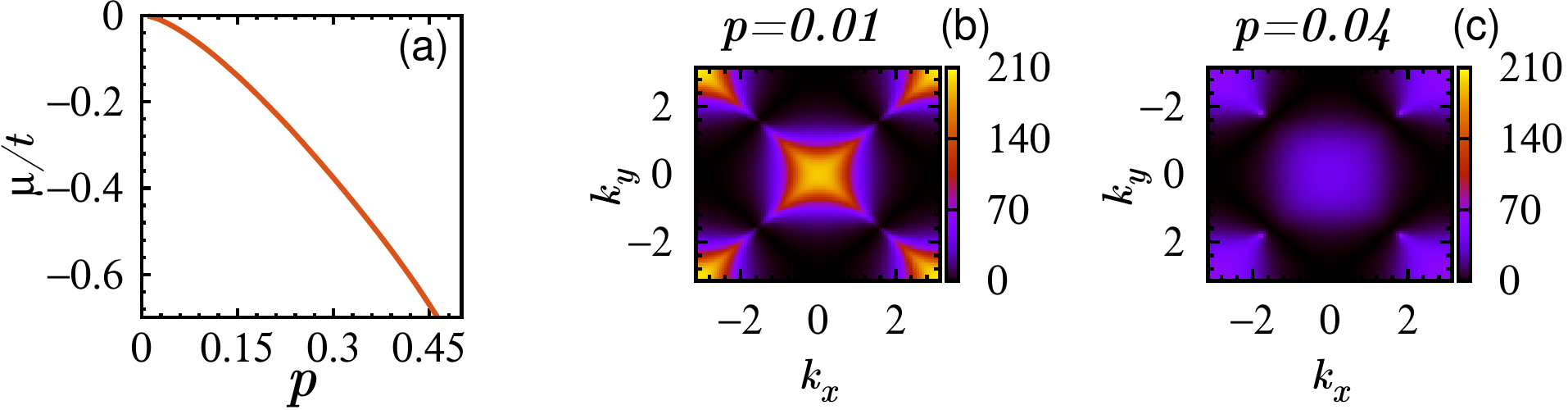}
		\caption{(color online) {\bf (a)} Chemical potential $\mu$ against doping $\doping$, {\bf (b)} the summand in the RHS of \eqn{eq:S1_MAPM} as a function of $k_x^{}$ and $k_y^{}$ at $\doping = 0.01$, and, {\bf (c)} the summand in the RHS of \eqn{eq:S1_MAPM} as a function of $k_x^{}$ and $k_y^{}$ at $\doping = 0.04$.}
		\label{fig:mu_MAPM}
	\end{figure}

	We note that the first term in the RHS of \eqn{eq:S1_MAPM} is positive, while the second term is negative. Furthermore, while the magnitude of the first term is inversely proportional to $J_P^{}(\doping) = \frac{g_s^{}(\doping) + g_n^{}(\doping)}{2}J$, the magnitude of the second term, in general, depends on the chemical potential $\mu(\doping)$, $d$-pairing scale $\Delta(\doping)$, and, most importantly, {\em the effective hopping amplitude $g_t^{}(\doping) t$}. It turns out, as is shown in the plots in \Fig{fig:mu_MAPM}, that at low doping ($\mu \sim 0$) the most important contribution to the second term comes from regions close to the nodal lines ($k_x^{} = k_y^{}$ and $k_x^{} = -k_y^{}$) where $\Delta_k^{} \sim 0$. That is, at low dopings the magnitude of the second term is largely determined by inverse of the effective hopping amplitude $g_t^{}(\doping) t$. Since $g_t(\doping) \rightarrow 0$ and $J_P(\doping) \approx J$ as $\doping \rightarrow 0$, it is clear that the negative term dominates at small doping indicating a critical $\doping_c$ at which $M_{P_a}$ vanishes.  A similar argument with \eqn{eq:S1_MAAM} also explains why $\MAAM$ would go to zero at some $\doping < \doping _c$. This is to be contrasted with $\MSAM$ (\eqn{eq:S1_MSAM}), which is {\em always} positive.
	
	This discussion highlights two points
\begin{enumerate}
 \item Only the antisymmetric modes are prone to being soft.
 \item The instability of the antisymmetric modes arises due to {\em strong correlation physics} that is encoded in the Gutzwiller factors. We emphasize that had it not been for the Gutzwiller factors, $\MAPM$ would not vanish for realistic values of $J$, i.~e., $J/t \sim 0.3$. 
\end{enumerate}

	To compute the stiffness ($\rho_s^{}$) corresponding to the Goldstone mode we integrate out all the massive modes to obtain a $\SPM$ only action in limit $q \rightarrow 0$
	\begin{equation}
		\label{eq:S1_SPM_action}
		\tilde{\action} \left[ \theta_s^{}(q \rightarrow 0) \right] \simeq \tilde{\action}^{SP}_{} + \sum_q \theta_{s}^*(q)\left\{ - \frac{\kappa}{2} \left( iq_l^{} \right) ^2_{} + \frac{\rho_s^{}}{2} |\mathbf{q} |^2_{} + O(q^4_{}) \right\} \theta_s^{}(q)
	\end{equation}
	where
	\begin{equation}
		\label{eq:S1_rho_s}
		\rho_{s}^{} (T=0) = \frac{\Delta^2_{}}{4\sites} \sum _\veck \left[ \frac{\left( \ckxmcky \right)^2_{}}{E_\veck^3} \left( |\boldsymbol{\nabla}_\veck^{} \tilde{\xi}_\veck^{}|^2_{} + |\boldsymbol{\nabla}_\veck^{} \Delta_\veck^{}|^2_{} \right) + \frac{\left( \cos 2k_x^{} + \cos 2k_y^{} - 2\ckxcky \right)}{E_\veck^{}} \right]
	\end{equation}
	and
	\begin{equation}
		\label{eq:S1_kappa}
		\kappa (T=0) = \frac{\Delta^2_{}}{2\sites} \sum _\veck \frac{\left( \ckxmcky \right)^2_{}}{E_\veck^3} - \frac{\left\{ \frac{1}{2\sites} \sum _\veck \tilde{\xi}_\veck^{} \frac{\left(\ckxmcky \right)^2_{}}{E_\veck^{3}}\right\}^2_{}}{\frac{1}{\sites} \sum_\veck \frac{\left(\ckxmcky \right)^4_{}}{E_\veck^{3}}}
	\end{equation}

\begin{figure}
		\centering
		\includegraphics[width=1.0\linewidth]{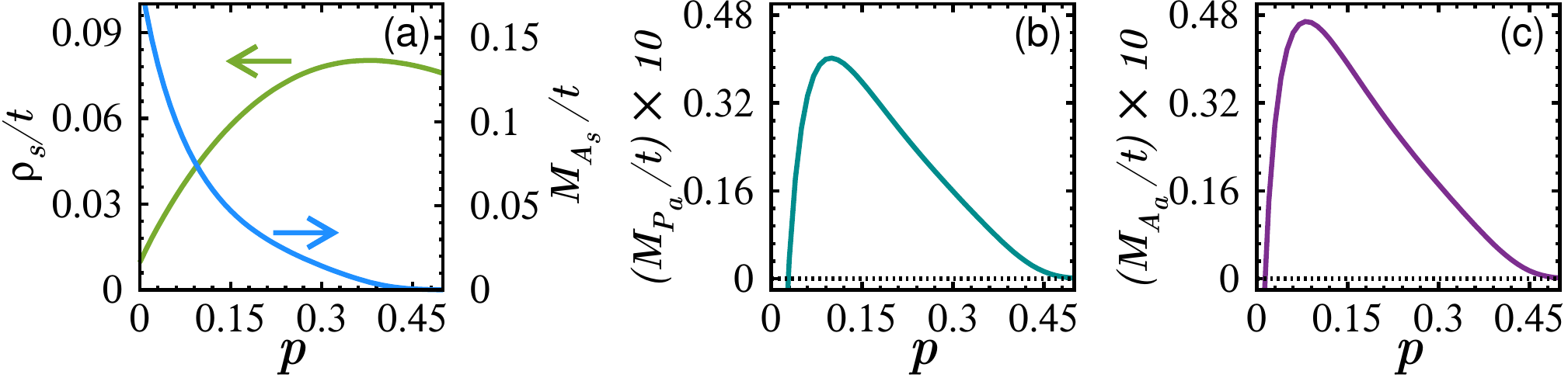}
		\caption{(color online) \textbf{Zhang {\em et al} Gutzwiller factors:} For $g_t^{}(\doping) = 2\doping / (1+\doping)$, $g_s^{}(\doping) = 4 / (1+\doping)^2_{}$ and $g_n^{}(\doping) = 1$ {\bf (a)} $\rho_s^{}/t$, the superfluid stiffness of the $\SPM$, and $\MSAM/t$, the gap parameter of the $\SAM$, {\bf (b)} $\MAPM/t$, the gap parameter of the $\APM$, and {\bf (c)} $\MAAM/t$, the gap parameter of the $\AAM$.}
		\label{fig:full_proj}
	\end{figure}

	In the main text of the paper, we have used $g_t(\doping) = \doping$, $g_s(p) = g_n(p) = 1$ to evaluate the gap parameters and the superfluid density. Here, to illustrate the point that the qualitative features of these calculations are independent of the choice of the Gutzwiller factors, we present in \Fig{fig:full_proj} the results obtained with a different choice of Gutzwiller factors, $g_t^{}(\doping) = 2\doping / (1+\doping)$, $g_s^{}(\doping) = 4 / (1+\doping)^2_{}$ and $g_n^{}(\doping) = 1$\cite{ShibaSST88}. The key point is that, while the quantitative aspects associated the critical value of $\doping_c$ are affected, we do find that the qualitative aspect associated with softness of the antisymmetric modes that renders the $d$-SC state fragile is faithfully reproduced (see \ref{Appen:large-N} for further discussion).

	\section{Saddle Point including fluctuations}
	\label{Appen:fluc_SP}
	To obtain the saddle point order parameters, apart from \eqn{eq:S1_SPeq_Delta} and \eqn{eq:S1_SPeq_K}, one also needs to fix the chemical potential $\mu$. The simplest way to do this is to use
	\be
	\label{eq:S2_mu_SP}
	\sites(1-p) = -\left( \frac{\partial}{\partial \mu} F^{SP}_{} \right) _{T=0}^{}
	\en
	where $F_{}^{SP}$ is the saddle point grand free energy (\eqn{eq:S1_S_SP}) without the contributions from fluctuations of the order parameters. A more accurate way to determine the saddle point would be to use $F_{}^{SP} + F_{}^{GF}$ in \eqn{eq:S2_mu_SP}, instead of just $F_{}^{SP}$, where $F_{}^{GF}$ is the Gaussian fluctuations contribution to the grand free energy (see, for example \cite{DienerPRA08}). In the analysis detailed in this section we compute $F_{}^{GF}$ arising from the Gaussian fluctuations of the $d$-pairing field and comment on the instability of the more accurate saddle point thus computed.

	This analysis can most conveniently be started from \eqn{eq:S1_action_gaussian}. The Gaussian contribution of the pair fluctuation fields ($\eta$) to the partition function is given by
	\begin{equation}
		\label{eq:S2_Z_eta_1}
		Z_\eta^{(2)} = \int \prod_q \text{d} \eta_0^*(q)\text{d} \eta_0^{}(q)\text{d} \eta_1^*(q)\text{d} \eta_1^{}(q)\ \exp{\left[ \Lambda \d (q) \mathcal{D} (q)^{-1}_{} \Lambda (q)\right]}
	\end{equation}
	With some algebra one can show that $\Lambda \d (-q) \mathcal{D} (-q)^{-1}_{} \Lambda (-q) = \Lambda \d (q) \mathcal{D} (q)^{-1}_{} \Lambda (q)$. This implies,
	\begin{equation}
		\label{eq:S2_Z_eta_2}
		Z_\eta^{(2)} = \int \prod_{q>0} \text{d} \eta_0^*(q)\text{d} \eta_0^{}(q)\text{d} \eta_1^*(q)\text{d} \eta_1^{}(q)\text{d} \eta_0^*(-q)\text{d} \eta_0^{}(-q)\text{d} \eta_1^*(-q)\text{d} \eta_1^{}(-q)\ \exp{\left[2 \Lambda \d (q) \mathcal{D} (q)^{-1}_{} \Lambda (q)\right]}
	\end{equation}
	Here, $q>0$ has the symbolic but well defined meaning that only half of the $q=(iq_l^{},\vecq)$ variables are involved in the evaluation of $Z_\eta^{(2)}$ in \eqn{eq:S2_Z_eta_2}. Now, using the Gaussian integration formula for complex variables
	\begin{equation}
		\label{eq:S2_Z_eta_3}
		Z_\eta^{(2)} = A \prod_{q>0} \left( \det{\mathcal{D} (q)^{-1}_{}} \right)^{-1}_{}
	\end{equation}
	\begin{equation}
		\label{eq:S2_Z_eta_3.1}
		= A\prod_{\vecq} \prod_{q_l^{}>0} \left( \det{\mathcal{D} (iq_l^{},\vecq)^{-1}_{}} \right)^{-1}_{} = A\prod_{\vecq} \prod_{q_l^{}>0} \left( \det{\mathcal{D} (-iq_l^{},\vecq)^{-1}_{}} \right)^{-1}_{}
	\end{equation}
	where $A$ is an unimportant constant. Again some algebra yields, $\det{\mathcal{D} (iq_l^{},\vecq)^{-1}_{}} = \left( \det{\mathcal{D} (-iq_l^{},\vecq)^{-1}_{}} \right)^*_{}$. This, along with \eqn{eq:S2_Z_eta_3.1}, proves that $Z_\eta^{(2)}$ is real. But, it may not be positive; a negative $Z_\eta^{(2)}$ would indicate that the system is unstable to fluctuations of the $d$-pairing field. Assuming that $Z_\eta^{(2)}$ evolves smoothly from being positive (system stable to $\eta$ fluctuations) to being negative (system unstable to $\eta$ fluctuations) as a function of hole doping $p$, and to visualize how the instability comes about through the long wavelength collective modes discussed in the main text we set
	\begin{equation}
		\label{eq:S2_Z_eta_4}
		Z_\eta^{(2)} = A\prod_{\vecq} \prod_{q_l^{}>0} \left( |\det{\mathcal{D} (iq_l^{},\vecq)^{-1}_{}} |\right)^{-1}_{}
	\end{equation}

	Then formally, apart for some unimportant constants
	\begin{equation}
		\label{eq:S2_F_GF_1}
		F_{}^{GF} = \frac{1}{\beta} \sum_{\vecq} \sum_{q_l^{}>0} \ln \left( |\det{\mathcal{D} (iq_l^{},\vecq)^{-1}_{}} |\right)
	\end{equation}
	The evaluation of the Matsubara sum in \eqn{eq:S2_F_GF_1} has to be done carefully. At the face of it, it appears to be non-convergent, but by tracking the convergence factors appearing because of time ordering of the path integral action one can get the correct $F_{}^{GF}$.
	\begin{equation}
		\label{eq:S2_F_GF_2}
		\begin{split}
			F_{}^{GF}(T=0) = \frac{1}{2\pi} \int_0^\infty \text{d}y \sum_\vecq^{} \left[ \ln \left( |\det{\mathcal{D} (iy^{},\vecq)^{-1}_{}} |\right) + 4\ln \left( 2J_P^{} \right) \right. & \left. - 4 J_P^{} \operatorname{Re} \left(C_{00}^{}(iy,\vecq) + C_{11}^{}(iy,\vecq) \right) \right]\\
			& + 2J_P^{} \sum_\vecq \left( \mathcal{C}_{00}^{}(\vecq) + \mathcal{C}_{11}^{}(\vecq) \right)
		\end{split}
	\end{equation}
	where
	\begin{equation}
		\label{eq:S2_curlyC}
		\mathcal{C}_{\alpha \gamma}^{} (\vecq )= \frac{1}{\beta} \sum_{q_l^{}} \text{e}^{iq_l^{}0^+_{}} C_{\alpha \gamma}^{} (iq_l^{},\vecq)
	\end{equation}

	Now, when we compute the more accurate saddle point as described in the first paragraph of this section, for several acceptable choices of the Gutzwiller factors\cite{ShibaSST88,AndersonJPCM04,LeeRMP06,BaskaranSSC87,Kotliar1PRB88}, we find that it is also unstable to $\APM$ fluctuations but at much higher values of hole doping ($0.12 \lesssim p_c^{} \lesssim 0.26$). While this, given the approximate way in which we implement projection, may not be accurate, it does indicate the possibility that cuprates may host an associated QCP in the doping range of interest.

	\section{Large-$\flavours$ formalism}
	\label{Appen:large-N}

	In this section we obtain the $1/\flavours$ corrections for the $\flavours$ flavor model introduced in the main text, $\delta \Delta$, $\delta K$ and $\delta \mu$, to the $\flavours \rightarrow \infty$ values of $\Delta$, $K$ and $\mu$, respectively, at $T=0$. In the large $\flavours$ limit the grand free energy density can be expanded in powers of $1/\flavours$ as (see, e.g., \cite{RadzihovskyPRA07})

	\be
	\label{eq:S3_e}
	\frac{1}{\flavours}\ \varepsilon (\mu,\Delta,K) = \varepsilon ^{(0)}_{} (\mu,\Delta,K) + \frac{1}{\flavours} \ \varepsilon ^{(1/\flavours)}_{} (\mu,\Delta,K) + \cdots
	\en
	where $\varepsilon^{(0)}_{} = F^{SP}_{}(T=0)/N$ of \eqn{eq:S1_S_SP} and

	\be
	\label{eq:S3_e-1/N}
	\varepsilon _{}^{(1/\flavours)}= -\frac{1}{(2\pi)^3_{}} \int _{0}^{\infty} dy \int _{BZ} d\vecq \quad \ln \frac{| \mathrm{Det} \mathcal{D}(iy,\vecq) |}{ (2 J_P^{})^4_{}}
	\en
	is the $1/\mathcal{N}$ correction to saddle point grand free energy density per $\lambda$ flavor at $T=0$ (cf $F^{GF}_{}/N$, \eqn{eq:S2_F_GF_1} and \ref{eq:S2_F_GF_2}, of $\flavours = 1$ theory).

In the large $\flavours$ limit, the chemical potential $\mu$, the uniform d-wave SC order parameter $\Delta$, and the uniform ``Fock'' parameter $K$ can be expanded in powers of $1/\flavours$ about the $\flavours \rightarrow \infty$ limit.

	\be
	\label{eq:S3_mu-1/N}
	\mu = \mu^{(0)}_{} + \frac{1}{\flavours}\ \delta \mu + \cdots
	\en

	\be
	\label{eq:S3_Delta-1/N}
	\Delta = \Delta^{(0)}_{} + \frac{1}{\flavours}\ \delta \Delta + \cdots
	\en

	\be
	\label{eq:S3_K-1/N}
	K = K^{(0)}_{} + \frac{1}{\flavours}\ \delta K + \cdots
	\en
	Note, $(\mu^{(0)}_{},\Delta^{(0)}_{},K^{(0)}_{})$ of the large-$\flavours$ theory is the same as the saddle point values $(\mu^{}_{SP},\Delta^{}_{SP},K^{}_{SP})$ of the $\flavours = 1$ theory. In particular, the following equations hold

	\be
	\label{eq:S3_e0/mu0}
	\left( \frac{\partial \varepsilon^{(0)}_{}}{\partial \mu}\right)_{\mu = \mu^{(0)}_{}}^{} = - (1-p)
	\en
	(cf \eqn{eq:S2_mu_SP})

	\be
	\label{eq:S3_e0/Delta0}
	\left( \frac{\partial \varepsilon^{(0)}_{}}{\partial \Delta}\right)_{\Delta = \Delta^{(0)}_{}}^{} = 0
	\en
	(cf \eqn{eq:S1_SP_Delta})

	\be
	\label{eq:S3_e0/K0}
	\left( \frac{\partial \varepsilon^{(0)}_{}}{\partial K}\right)_{K = K^{(0)}_{}}^{} = 0
	\en
	(cf \eqn{eq:S1_SP_K})

	Now, using \eqn{eq:S3_mu-1/N}-\ref{eq:S3_K-1/N} in \eqn{eq:S3_e}, and setting $(\partial _{\mu}^{},\partial _{\Delta}^{},\partial _{K}^{}) \varepsilon = (p-1,0,0)$, to order $1/\flavours$, one obtains

	\begin{equation}
	\label{eq:S3_large-N_correction}
	\begin{pmatrix}
	\delta \mu \\
	\delta \Delta \\
	\delta K \\
	\end{pmatrix}
	=
	- \begin{pmatrix}
	\partial_{\mu \mu}^{} \varepsilon_{}^{(0)} & \partial_{\mu \Delta}^{} \varepsilon_{}^{(0)} & \partial_{\mu K}^{} \varepsilon_{}^{(0)} \\
	\partial_{\Delta \mu}^{} \varepsilon_{}^{(0)} & \partial_{\Delta \Delta}^{} \varepsilon_{}^{(0)} & \partial_{\Delta K}^{} \varepsilon_{}^{(0)} \\
	\partial_{K \mu}^{} \varepsilon_{}^{(0)} & \partial_{K \Delta}^{} \varepsilon_{}^{(0)} & \partial_{K K}^{} \varepsilon_{}^{(0)} \\
	\end{pmatrix}^{-1}_{}
	\begin{pmatrix}
	\partial _\mu^{} \varepsilon _{}^{(1)} \\
	\partial _\Delta^{} \varepsilon _{}^{(1)} \\
	\partial _K^{} \varepsilon _{}^{(1)} \\
	\end{pmatrix}
	\end{equation}
for the $1/\mathcal{N}$ corrections, $\delta \mu$, $\delta \Delta$ and $\delta K$, to the $\flavours \rightarrow \infty$ (saddle point of $\flavours = 1$) values of $\mu$, $\Delta$ and $K$, respectively. Note that, all the partial derivatives in \eqn{eq:S3_large-N_correction} have to be evaluated at $\flavours \rightarrow \infty$ values of $\mu$, $\Delta$ and $K$.

\begin{figure}
		\centering
		\includegraphics[width=0.5\linewidth]{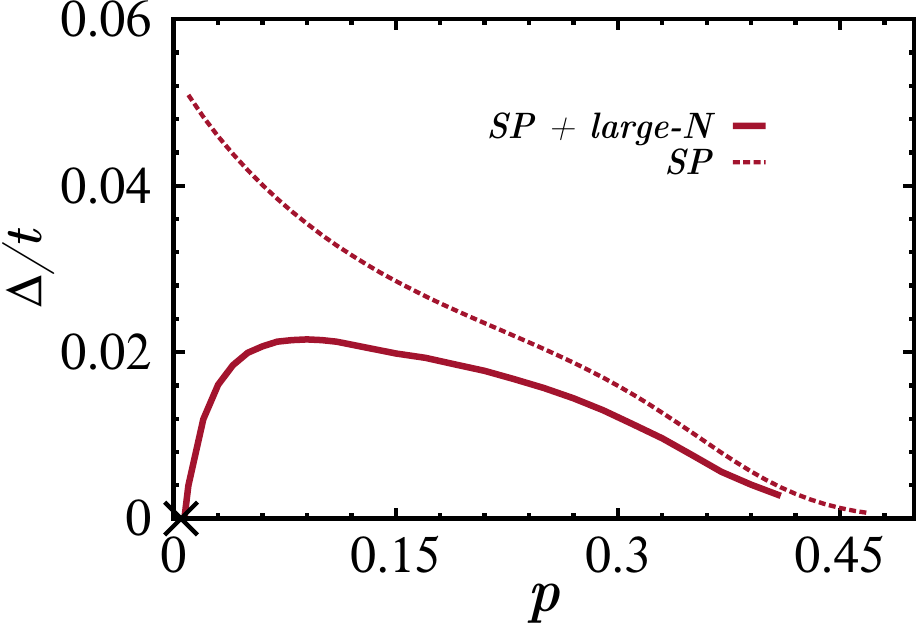}
		\caption{(color online) \textbf{Large-N correction with Lee {\em et al} Gutzwiller factors:} For $g_t^{}(\doping) = \doping$, $g_s^{}(\doping) = 3/2$ and $g_n^{}(\doping) = 0$, $d$-pairing scale obtained from large-$\flavours$ theory is shown by the solid red curve. The saddle point value of the scale is shown, for reference, by the dashed red line. The cross on the doping axis marks the $\APM$ instability of the saddle point.}
		\label{fig:SBMFT_large-N}
	\end{figure}

To further illustrate the point that the results presented in the main text are largely independent of the choice of the Gutzwiller factors, we now present in \Fig{fig:SBMFT_large-N} the results for the $d$-pairing scale within the large-$\flavours$ theory with $g_t^{}(\doping) = \doping$, $g_s^{}(\doping) = 3/2$ and $g_n^{}(\doping) = 0$\cite{LeeRMP06}. Clearly, this again displays the very same qualitative aspects associated with the softness of the antisymmetric modes found by a different choice of the Gutzwiller factors in the main text of the paper.

	\section{Ginzburg-Landau Functional}
	\label{Appen:GL}
	In this section we derive a Landau functional from \ref{eq:S1_action_bosonic}, using $\Delta_\alpha(q) = \sqrt{\sites \beta}\Delta_\alpha \delta_{q,0}$, $K_\alpha^{}(q) =0$, and assuming $\Delta_\alpha^{}$ to be small. We show that it has all the required symmetries. We then extend it to include the $q\neq0$ fluctuations\cite{BerlinskyPRB94,BerlinskyPRL95}, and study the nature of long wavelength collective modes around the $d$-wave state. We also study the coupling of the collective phase modes with the electromagnetic gauge field.

	Using $\Delta_\alpha(q) = \sqrt{\sites \beta} \Delta_\alpha \delta_{q,0}$ and $K_\alpha^{}(q) =0$, \ref{eq:S1_action_bosonic} becomes
	\begin{equation}
		\label{eq:S4_act_bos_K0_q0}
		\tilde{\action} \left[ \{\Delta_{\alpha'}^{}\} \right]= \frac{2 \sites \beta}{J_P^{}} \left( |\Delta_x ^{} |^2_{} + |\Delta_x ^{} |^2_{}\right) - \sum_k \ln \left[ (ik_n^{})^2_{} - \xi_\veck^{2} -2 |\Delta_x ^{}\cos k_x^{} + \Delta_y ^{}\cos k_y^{}|^2_{} \right]
	\end{equation}
	Note, $\Delta_\alpha(q) = \sqrt{\sites \beta} \Delta_\alpha \delta_{q,0}$ implies $\Delta_\alpha(\tau, \mathbf{r}) = \Delta_\alpha$. Defining
	\begin{equation}
		\label{eq:S4_sd}
		s(\tau, \vecr) = \left( \Delta_x^{}(\tau, \vecr) + \Delta_y^{}(\tau, \vecr)\right)/2 \qquad \text{and} \qquad d(\tau, \vecr) = \left( \Delta_x^{}(\tau, \vecr) - \Delta_y^{}(\tau, \vecr)\right)/2
	\end{equation}
	as the extended $s$ and $d$-wave pairing amplitudes, respectively, we obtain
	\begin{equation}
		\label{eq:S4_act_sd}
		\frac{\tilde{\action} \left[ \{s,d\} \right]}{\sites \beta}= \frac{2}{J_P^{}} \left( |s|^2_{} + |d|^2_{}\right) - \frac{1}{\sites \beta} \sum_k \ln \left[ (ik_n^{})^2_{} - \xi_\veck^{2} \right] - \frac{1}{\sites \beta} \sum_k \left[ 1 - \frac{2 |(s+d)\cos k_x^{} + (s-d)\cos k_y^{}|^2_{}}{(ik_n^{})^2_{} - \xi_\veck^{2}} \right]
	\end{equation}
	Now, assuming $s$ and $d$ to be small and setting $1/\beta = T = 0$, we get
	\begin{equation}
		\label{eq:S4_f_L}
		f_L^{} = \alpha_s^{} |s|^2_{} + \alpha_d^{} |d|^2_{} + \beta_1^{} |s|^4_{} + \beta_2^{} |d|^4_{} + \beta_3^{} |s|^2_{} |d|^2_{} + \beta_4^{} \left( s^{*2}_{} d^2_{} + d^{*2}_{} s^2_{} \right)
	\end{equation}
	where we have dropped some constant terms, and
	\begin{equation}
		\label{eq:S4_alpha_s}
		\alpha_s^{} = \frac{2}{J_P^{}} - \frac{1}{\sites} \sum_{\stackrel{\veck}{\xi_\veck^{} > \Delta_\veck^{}\sim 0}} \frac{(\cos k_x^{}+\cos k_y^{})^2_{}}{|\xi_\veck^{}|} \sim \frac{2}{J_P^{}} - \frac{1}{\sites} \sum_{\veck} \frac{(\cos k_x^{}+\cos k_y^{})^2_{}}{E_\veck^{}}
	\end{equation}
	\begin{equation}
		\label{eq:S4_alpha_d}
		\alpha_d^{} = \frac{2}{J_P^{}} - \frac{1}{\sites} \sum_{\stackrel{\veck}{\xi_\veck^{} > \Delta_\veck^{}\sim 0}} \frac{(\cos k_x^{}-\cos k_y^{})^2_{}}{|\xi_\veck^{}|} \sim \frac{2}{J_P^{}} - \frac{1}{\sites} \sum_{\veck} \frac{(\cos k_x^{}-\cos k_y^{})^2_{}}{E_\veck^{}}
	\end{equation}
	\begin{equation}
		\label{eq:S4_beta_12}
		\beta_1^{} = \frac{1}{2 \sites} \sum_{\stackrel{\veck}{\xi_\veck^{} > \Delta_\veck^{}\sim 0}} \frac{(\cos k_x^{}+\cos k_y^{})^4_{}}{|\xi_\veck^{}|^3_{}}, \qquad \beta_2^{} = \frac{1}{2 \sites} \sum_{\stackrel{\veck}{\xi_\veck^{} > \Delta_\veck^{}\sim 0}} \frac{(\cos k_x^{}-\cos k_y^{})^4_{}}{|\xi_\veck^{}|^3_{}}
	\end{equation}
	\begin{equation}
		\label{eq:S4_beta_3}
		\beta_3^{} = \frac{2}{\sites} \sum_{\stackrel{\veck}{\xi_\veck^{} > \Delta_\veck^{}\sim 0}} \frac{(\cos^2_{} k_x^{}-\cos^2_{} k_y^{})^2_{}}{|\xi_\veck^{}|^3_{}}, \qquad \beta_4^{} = \frac{\beta_3^{}}{4}.
	\end{equation}
	Clearly, $f_L^{}$ is consistent with all the symmetries of the original Hamiltonian. Two more points about $f_L^{}$ are worth noting. First, for overdoped system, where $\Delta_\alpha^{}$ is small, $\alpha _d^{} < 0$ and all other parameters are positive (this is obvious for the $\beta$'s and can be shown numerically for the $\alpha$'s at $\doping=0.59$ with the Gutzwiller factors used in the main text). This is consistent with our saddle point theory where we find $d$-wave pairing to be stable for the overdoped case. Second, positive value of $\beta_4^{}$ indicates that fluctuations towards a $d+is$ state are favored compared to those towards a $d+s$ state (seen by setting $s=|s|\exp(i \phi _s^{})$ and $d=|d|$ in \eqn{eq:S4_f_L}).

	We now extend our Landau functional to a Ginzburg-Landau functional which includes spatial fluctuations and the electromagnetic gauge field\cite{BerlinskyPRB94,BerlinskyPRL95}
	\begin{equation}
		\label{eq:S4_f_GL}
		\begin{split}
		f_{GL} = \alpha_s^{} |s|^2_{} + \alpha_d^{} |d|^2_{} + \beta_1^{} |s|^4_{} + \beta_2^{} |d|^4_{} + \beta_3^{} |s|^2_{} |d|^2_{} + \beta_4^{} \left( s^{*2}_{} d^2_{} + d^{*2}_{} s^2_{} \right) + \gamma_s^{} |\mathbf{D} s|^2_{} + \gamma_d^{} |\mathbf{D} d|^2_{}\\
		+ \gamma_{\nu}^{} \{ \left( D_y^{} s \right)^*_{} \left( D_y^{} d \right) - \left( D_x^{} s \right)^*_{} \left( D_x^{} d \right) + \text{c.c.} \}
		\end{split}
	\end{equation}
	where
	\begin{equation}
		\label{eq:S4_D}
		\mathbf{D} = -i\boldsymbol{\nabla} - 2e \mathbf{A}
	\end{equation}
	with $\mathbf{A}$ being the electromagnetic vector potential. Note, $\gamma_s^{}$ and $\gamma_d^{}$ are expected to be positive. In the absence of $\mathbf{A}$, this leads to an uniform $d$-wave SC order with (remember, $\alpha_d^{}<0$ and $\alpha_s^{}>0$)
	\begin{equation}
		\label{eq:S4_Delta_SP}
		d(\vecr) = \left( -\frac{\alpha_d^{}}{2\beta_2^{}} \right)^{1/2} \equiv \Delta
	\end{equation}
	Next we set
	\begin{equation}
		\label{eq:S4_Delta_x}
		\Delta_x^{}(\vecr) = \Delta \left( 1 + \zeta_x^{}(\vecr) \right)\text{e}^{i\theta_x^{}(\vecr)}_{}
	\end{equation}
	and
	\begin{equation}
		\label{eq:S4_Delta_y}
		\Delta_y^{}(\vecr) = -\Delta \left( 1 + \zeta_y^{}(\vecr) \right)\text{e}^{i\theta_y^{}(\vecr)}_{}
	\end{equation}
	in \eqn{eq:S4_f_GL} and obtain $\MSAM$, $\MAAM$ and $\MAPM$ as the coefficients of $\frac{1}{2}\left( \frac{\zeta_x^{} + \zeta_y^{}}{2} \right)^2_{}$, $\frac{1}{2}\left( \frac{\zeta_x^{} - \zeta_y^{}}{2} \right)^2_{}$ and $\frac{1}{2}\left( \frac{\theta_x^{} - \theta_y^{}}{2} \right)^2_{}$, respectively.
	\begin{equation}
		\label{eq:S4_MSAM}
		\MSAM = 8\beta_2^{}\Delta^4_{}
	\end{equation}
	\begin{equation}
		\label{eq:S4_MAAM}
		\MAAM = 2\alpha_s^{}\Delta^2_{} + 2\left( \beta_3^{}+2\beta_4^{}\right)\Delta^4_{} = 2\alpha_s^{}\Delta^2_{} + 3 \beta_3^{}\Delta^4_{}
	\end{equation}
	and
	\begin{equation}
		\label{eq:S4_MAPM}
		\MAPM = 2 \alpha_s^{} \Delta^2_{} + 2\left( \beta_3^{} - 2\beta_4^{}\right) \Delta^4_{} = 2\alpha_s^{}\Delta^2_{} + \beta_3^{}\Delta^4_{}
	\end{equation}
	All the gap parameters are positive, which is concordant with our saddle point results in the overdoped region. We also find that the coefficient of $\frac{1}{2}\left( \frac{\theta_x^{} + \theta_y^{}}{2} \right)^2_{}$ is zero, establishing it as the Goldstone mode.

	To study the coupling of the electromagnetic gauge field with the phase fluctuation modes we set $\zeta_x^{}$ and $\zeta_y^{}$ to zero to obtain
	\begin{equation}
		\label{eq:S4_f_eff}
		\begin{split}
		f \simeq f^{MF}_{} + \left\{ \MAPM + 2(\gamma_s^{}-\gamma_d^{} ) \Delta^2_{} |\boldsymbol{\nabla} \theta_s^{} - 2e\mathbf{A}|^2_{}\right\} \frac{\theta_a^2}{2} + \gamma_d^{} \Delta_{}^2 |\boldsymbol{\nabla} \theta_s^{} - 2e\mathbf{A}|^2_{} + \gamma_s^{} \Delta_{}^2 |\boldsymbol{\nabla} \theta_a^{} |^2_{} \\
		+ 2\gamma_{\nu}^{} \Delta_{}^2 \left\{ \partial_y^{} \theta_a^{} ( \partial_y^{} \theta_s^{} - 2eA_y^{}) - \partial_x^{} \theta_a^{} ( \partial_x^{} \theta_s^{} - 2eA_x^{}) \right\}
		\end{split}
	\end{equation}
	where $\theta_s^{} = \frac{\theta_x^{} + \theta_y^{}}{2}$, $\theta_a^{} = \frac{\theta_x^{} - \theta_y^{}}{2}$ and
	\begin{equation}
		\label{eq:S4_f_MF}
		f_{}^{MF} = \alpha_d^{} \Delta^2_{} + \beta_2^{} \Delta^4_{} = -\frac{\alpha^2_d}{4\beta_2^{}}
	\end{equation}
	The gauge invariant charge current is given by
	\begin{equation}
		\label{eq:S4_j}
		\mathbf{j} \simeq 4e\gamma_d^{} \Delta^2_{} (2e\mathbf{A} - \boldsymbol{\nabla} \theta_s^{}).
	\end{equation}
	Therefore, \eqn{eq:S4_f_eff} and \eqn{eq:S4_j} imply
	\begin{equation}
		\label{eq:S4_MAPM_eff}
		\MAPM^{eff} = \MAPM + 2(\gamma_s^{}-\gamma_d^{} ) \left( \frac{j}{4e\gamma_d^{} \Delta} \right) ^2_{}
	\end{equation}
	The effective gap parameter of $\APM$ is sensitive to the presence of charge currents in the system. Note that the symmetric phase mode couples to the electromagnetic field in the standard gauge invariant way (third term in \eqn{eq:S4_f_eff}). As is well known\cite{AltlandCUP06}, it can be integrated out to generate a gauge invariant mass term for the transverse component of the electromagnetic field leading to the Meissner effect. In other words, the symmetric phase mode fluctuations, which are conjugate to charge fluctuations, couple with the electromagnetic field to give rise to a collective mode with mass pushed up to the plasma frequency.



\end{document}